\documentstyle[honnef,pslatex]{article}
\newcommand{\J}{J/\psi}
\newcommand{\PRL}{Phys. Rev. Lett.}
\newcommand{\PL}{Phys. Lett.}
\newcommand{\JPG}{J. Phys. G}
\newcommand{\NP}{Nucl. Phys.}
\newcommand{\etal}{et al.}
\frompage{000} \topage{000}
\def\rightmark{Current Status of Quark Gluon Plasma Signals} 
\def\leftmark{D. Zschiesche et al.}

\title{Current Status of Quark Gluon Plasma Signals}

\authors{%
{D.~Zschiesche$^1$,
S.~Bass$^{2,3}$,
M.~Bleicher$^1$,
J.~Brachmann$^1$, 
L.~Gerland$^1$, 
K.~Paech$^1$,
S.~Scherer$^1$, 
S.~Soff$^4$,
C.~Spieles$^1$,
H.~Weber$^1$, 
H.~St\"ocker$^1$, 
W.~Greiner$^1$
}
\\[2.812mm]
{\normalsize
\hspace*{-8pt}$^1$ Institut f\"ur Theoretische Physik, J.W. Goethe
Universit\"at  \\ 
60054 Frankfurt a.M., Germany\\[0.2ex] 
\hspace*{-8pt}$^2$ Department of Physics, Duke University\\ 
27708-0305 Durham, NC, USA\\[0.2ex]
\hspace*{-8pt}$^3$ RIKEN BNL Research Center, Brookhaven National Laboratory\\ 
Upton, NY 11973, USA\\[0.2ex]
\hspace*{-8pt}$^4$ Physics Department, Brookhaven National
Laboratory\\ 
Upton, NY 11973, USA\\ 
}}
\abstract{
Compelling evidence for the creation of a new form 
of matter has been claimed to be found in 
Pb+Pb collisions at SPS.
We discuss 
the uniqueness of often proposed experimental signatures 
for quark matter formation in relativistic heavy ion collisions. 
It is demonstrated that so far none of the proposed signals like
$J/\psi$ meson production/suppression, strangeness enhancement,   
dileptons, and directed flow unambigiously show that a phase of
deconfined matter has been formed in SPS Pb+Pb collisions.
We emphasize the need for systematic future measurements 
to search for simultaneous irregularities in the excitation functions 
of several observables in order to come close to pinning 
the properties of hot, dense QCD matter from data.
}
\begin{document}
\maketitle
\section{Introduction}
In the last few years researchers at Brookhaven and CERN have 
succeeded to measure a wide spectrum of observables with 
heavy ion beams, $Au+Au$ and $Pb+Pb$. While these programs 
continue to measure with greater precision the beam energy-, 
nuclear size-, and centrality dependence of those observables, 
it is important to recognize the major 
milestones passed thusfar in that work. 
The experiments have conclusively demonstrated the existence 
of strong nuclear $A$ dependence of, among others, $J/\psi$ 
and $\psi'$ meson production and suppression, strangeness enhancement, 
hadronic resonance production, 
stopping and directed collective transverse and longitudinal 
flow of baryons and mesons -- in and out of the impact plane, 
both at AGS and SPS energies --, 
and dilepton-enhancement below and above the $\rho$ meson mass. 
These observations support that a novel form of ``resonance matter'' 
at high energy- and baryon density has been created in nuclear collisions.
The global multiplicity and transverse energy measurements prove 
that substantially more entropy is produced in $A+A$ collisions at the SPS
than simple superposition of $A\times pp$ would imply. 
Multiple initial and final state interactions play a critical role 
in all observables. 
The high midrapidity baryon density (stopping) and the observed 
collective transverse and directed flow patterns constitute one
of the strongest evidence for the existence of an extended period 
($\Delta \tau\approx 10$~fm/c) of high pressure and 
strong final state interactions. The enhanced $\psi'$ 
suppression in $S+U$ relative to $p+A$ also attests to this fact. 
The anomalous low mass dilepton enhancement shows that substantial 
in-medium modifications of multiple collision dynamics exist, 
probably related to in-medium collisional broadening of vector mesons. 
The non-saturation of the strangeness (and anti-strangeness) 
production shows that novel non-equilibrium production 
processes arise in these reactions. 
Finally, the centrality dependence of $J/\psi$ absorption 
in $Pb+Pb$ collisions presents further hints towards the nonequilibrium 
nature of such reactions. 
Is there evidence for the long sought-after quark-gluon plasma 
that thusfar has only existed as a binary array of predictions 
inside teraflop computers?

As we will discuss, it is too early to tell. 
Theoretically there are still too many ``scenarios'' and 
idealizations to provide a satisfactory answer. 
Recent results from microscopic transport models as well as 
macroscopic hydrodynamical calculations differ significantly 
from predictions of simple thermal models, e.~g.\ in the flow pattern. 
Still, these nonequilibrium models provide reasonable predictions 
for the experimental data. We may therefore be forced to rethink our
concept of what constitutes the deconfined phase in ultrarelativistic 
heavy-ion collisions. Most probably it is not a blob of thermalized quarks and
gluons. Hence, a quark-gluon plasma can only be the 
source of {\em differences} to the predictions of these models for 
hadron ratios, the $J/\psi$ meson production, dilepton yields, 
or the excitation function of transverse flow.
And there are experimental gaps such as the lack of 
intermediate mass $A\approx 100$ data and the limited number of beam energies 
studied thusfar, in particular between the AGS and SPS. 
Now the field is at the doorstep of the next milestone: 
$A+A$ at $\sqrt{s}=30-200$ $\rm{AGeV}$ which have started a few months ago.
\section{J/$\psi$ production}
The QCD factorization theorem is used to evaluate the PQCD cross sections of
heavy quarkonium interactions with ordinary hadrons. However, the charmonium
states (here denoted $X$) are not sufficiently small to ignore nonperturbative
QCD physics.
Thus, we evaluate the nonperturbative QCD contribution to the cross sections
of charmonium-nucleon interaction by using an interpolation between
known cross sections~\cite{ger}. 
The $\J$-$N$ cross section evaluated in this paper is
in reasonable agreement with SLAC data \cite{slac}.

Indeed, the $A$-dependence of the $\J$ production studied at SLAC
at $E_{inc} \sim 20$ $\rm{GeV}$ exhibits a significant absorption effect \cite{slac}
leading to $\sigma_{abs}(\J$-$N)= 3.5 \pm 0.8$ mb.
It was demonstrated~\cite{farrar} that, in the kinematic region at SLAC,
the color coherence effects are still small on the internucleon
scale for the formation of $\J$'s.
So, in contrast to the findings at higher energies, at intermediate
energies this process measures the {\it genuine} $\J$-$N$
interaction cross section at energies of $\sim $ 15-20 $\rm{GeV}$ \cite{farrar}.

To evaluate the nonperturbative QCD contribution we use an
interpolation formula~\cite{ger} for the dependence of the cross section
on the transverse size $b$ of a quark-gluon configuration
Three reference points are used to fix our parametrization of
the cross sections (cf. Tab.~\ref{meanb}).
The $X$-$N$ cross sections is calculated via:
$
\sigma=\int \sigma(b)\cdot |\Psi (x,y,z)|^2 {\rm d}x\, {\rm d}y\,
{\rm d}z 
$,
where $\Psi (x,y,z)$ is the charmonium wave function. In our calculations
we use the wave functions from a non-relativistic charmonium model 
(see~\cite{werner}).%

\begin{table}[hb]
\vspace*{-18pt}
\caption
{\label{meanb}\small{The total quarkonium-nucleon cross sections $\sigma$.
For the $\chi$ two values arise, due to the spin dependent wave functions
($lm=10,11$).}}
\vspace*{-14pt}
\begin{center}
\begin{tabular}{|c|c|c|c|c|} 
\hline
$c\overline{c}$-state & J/$\psi$ & $\psi'$ & $\chi_{c10}$ & $\chi_{c11}$\\
\hline
$\sigma$ (mb) & 3.62 & 20.0 & 6.82 & 15.9 \\
\hline
\end{tabular}
\end{center} 
\end{table}
\vspace{-0.6cm}

We follow the analysis of~\cite{kharzeev} to evaluate
the fraction of $\J$'s (in $pp$ collisions) that come from the decays of
the $\chi$ and $\psi'$. So, the
suppression factor $S$ of $\J$'s produced in the nuclear medium is
calculated as:\\
$
S=0.6\cdot ( 0.92\cdot S^{\J}+0.08\cdot S^{\psi'})+0.4\cdot S^{\chi}
$.
Here $S^X$ are the respective suppression factors of the different
pure charmonium states $X$ in nuclear
matter. The $S^X$ are for minimum bias $pA$ collisions within the
semiclassical approximation (cf.~\cite{hufner}).

The charmonium states are produced as small
configurations, then they evolve to
their full size.
Therefore, if the formation length of the charmonium states, $l_f$, becomes
larger than the average internucleon distance,
 one has to take into account the evolution of
the cross sections with the distance from the production point~\cite{farrar}.

The formation length of the $\J$ is given by 
$l_f\approx {2p}/{(m^2_{\psi'}- m^2_{\J})}$, where $p$ is the momentum of
the $\J$ in the rest frame of the target. For a $\J$ produced at midrapidity 
at SPS energies, this yields $l_f\approx 3$ fm.
Due to the lack of better knowledge, we use the same $l_f\approx 3$ fm for the
$\chi$. For the $\psi'$ we use $l_f\approx 6$ fm, 
because it is not a small object, but has the size of
a normal hadron, i.e. the pion. For $E_{lab}=800$ $\rm{AGeV}$ we get a factor
of two for the formation lengths due to the larger Lorentz factor.

\begin{figure}[htp]
\vspace*{-0.4cm}
%\insertplot{fig1.ps}
\centerline{
\parbox[b]{6cm}{\epsfxsize=6.5cm
\epsfbox{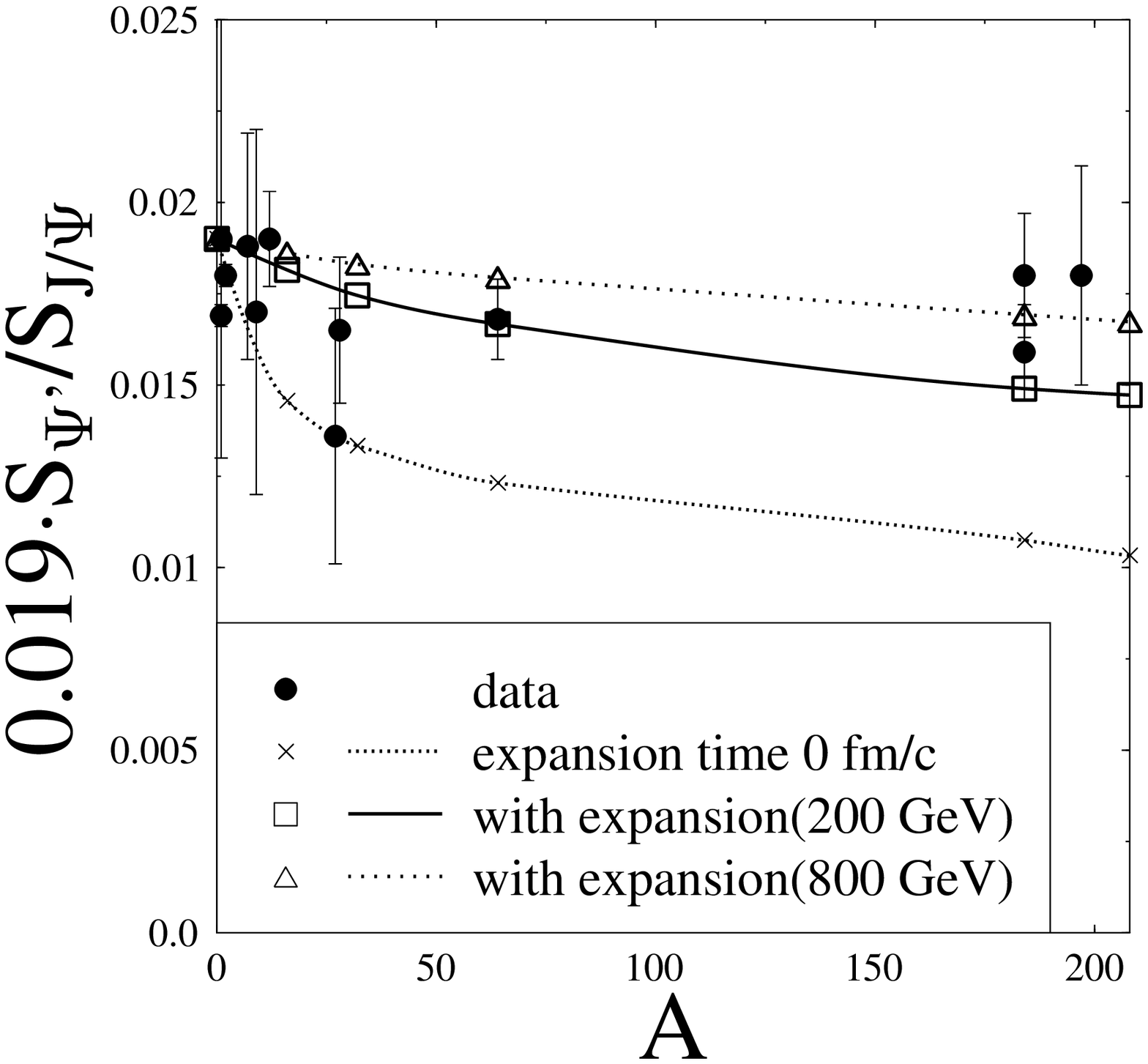}}
\hfill
\parbox[b]{6cm}{\epsfxsize=6.5cm
\epsfbox{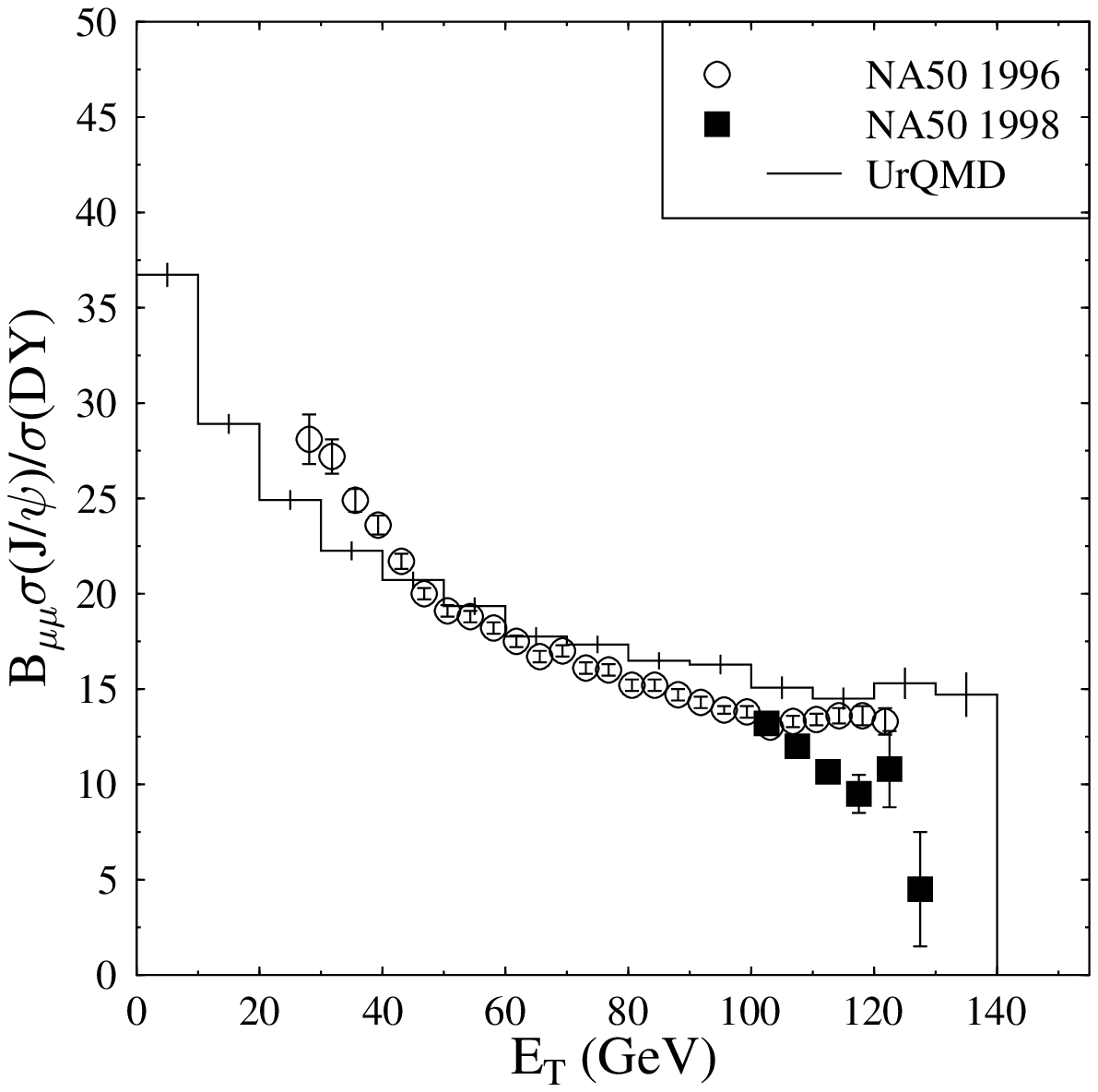}}
\hfill
\vspace*{-1.3cm} 
}
\caption
{Left: The ratio $0.019\cdot S_{\psi'} / S_{\J}$ is shown in $pA$ (crosses)
in comparison to the data (circles). The squares and the triangles shows the
ratio calculated with the expansion of small wave packages. 
Right: 
The ratio of $J/\psi$ to Drell-Yan production as a function of
$E_T$ for Pb+Pb at 160~$\rm{GeV}$. \label{psidyet}}
\vspace*{-0.9cm}
\end{figure}

However, this has a large impact on the
$\psi'$ to $\J$-ratio depicted in Fig.~\ref{psidyet}, which
shows the ratio $0.019\cdot S_{\psi'} / S_{\J}$ calculated with (squares
(200 $\rm{GeV}$) and triangles (800 $\rm{GeV}$))
and without (crosses) expansion. The factor 0.019 is the measured value in
$pp$ collisions, because the experiments do not measure the calculated value
$S_{\psi'} / S_{\J}$ but
${(B_{\mu\mu}\sigma(\psi'))}/{(B_{\mu\mu}\sigma(\J))}$.
$B_{\mu\mu}$ are the branching ratios for $\J ,\,\psi'\rightarrow\mu\mu$.

The calculations which take into
account the expansion of small wave packages show better agreement
with the data (circles) (taken from~\cite{carlos}) than the calculation
without expansion time, i.e. with immediate $\J$ formation, $l_f=0$.   
We calculated this effect both at $E_{lab}=200 \rm{AGeV}$ and 800 
$\rm{AGeV}$.     
The data have been measured at different energies ($E_{lab}$ = 200, 300, 400,
450, 800 $\rm{GeV}$ and $\sqrt{s} = 63 \rm{GeV}$). One can see that this ratio is nearly
constant in the kinematical region of the data, but it decreases at smaller
momentum (e.g. $E_{lab}=200 \rm{AGeV}$ and $y<0$) due to the larger cross section 
of the $\psi'$.

However, the P-states yield two vastly different
cross sections (see Tab.~\ref{meanb}) for $\chi_{10}$ and $\chi_{11}$,
respectively. This leads to a higher absorption rate of the $\chi_{11}$ as
compared to the $\chi_{10}$. This new form of color filtering is predicted
also for the corresponding states of other hadrons; e.g. for the bottomium
states which are proposed as contrast signals to the $\J$'s at RHIC and LHC!

Furthermore it is important to also take into account
comoving mesons. 
Therefore we use the UrQMD model \cite{bass98,spieles99}. 
Particles
produced by string fragmentation are not allowed to interact with other
 hadrons --
in particular with a charmonium state -- within their formation time 
(on average, $\tau_F\approx 1$ fm/c). However,
leading hadrons are allowed to interact with a reduced cross section even
within their formation time . 
The reduction factor is 1/2 for mesons which
contain a leading constituent quark from an incident nucleon and 2/3 for 
baryons which contain a leading diquark.

Figure~\ref{psidyet} shows the $J/\psi$ to Drell-Yan ratio
as a function of $E_T$ for Pb+Pb interactions at 160~$\rm{GeV}$ compared to
the NA50 data \cite{romana,na50neu}.
The normalization of $B_{\mu\mu}\sigma(J/\psi)/\sigma({\rm DY})=46$ in $pp$
interactions at 200~$\rm{GeV}$ has been
fit  to S+U data within a geometrical
model \protect\cite{kharzeev}.

The application of this value to our analysis is not arbitrary:
the model of Ref.~\protect\cite{kharzeev} renders
the identical $E_T$-integrated $J/\psi$ survival probability, $S=0.49$,
as the UrQMD calculation for this system.
An additional factor of 1.25
\protect\cite{reviewvogt} has been applied to the Pb+Pb calculation
in order to account for the lower energy, 160 $\rm{GeV}$, since the
$J/\psi$ and Drell-Yan cross sections have different energy and isospin
dependencies.

The gross features of the $E_T$ dependence of the $J/\psi$ to Drell-Yan
ratio are reasonably well described by the model calculation.
No discontinuities in the shape of the ratio as a function of $E_T$
are predicted by the
simulation. The new high $E_T$ data \cite{na50neu} decreases stronger
than the calculation. This could be caused by underestimated 
fluctuations of the 
multiplicity of secondaries in the UrQMD model. This occurs, since
high $E_T$-values are a trigger for very central events with a
secondary multiplicity larger than in average \cite{cap00}. 
\section{Dilepton production}

Beside results from hadronic probes, electromagnetic radiation 
-- and in particular dileptons -- 
offer an unique probe from the hot and dense reaction zone: 
here, hadronic matter is almost transparent. 
The observed enhancement of the dilepton yield 
at intermediate invariant masses ($M_{e^+e^-} > 0.3~\rm{GeV}$) 
received great interest: it was prematurely thought that
the lowering of vector meson masses is required by chiral symmetry restoration 
(see e.g.\cite{koch97} for a review). 
\begin{figure}[h] 
\vspace{-0.5cm} 
\centerline{\parbox[b]{6cm}
{\epsfxsize=6.5cm
\epsfbox{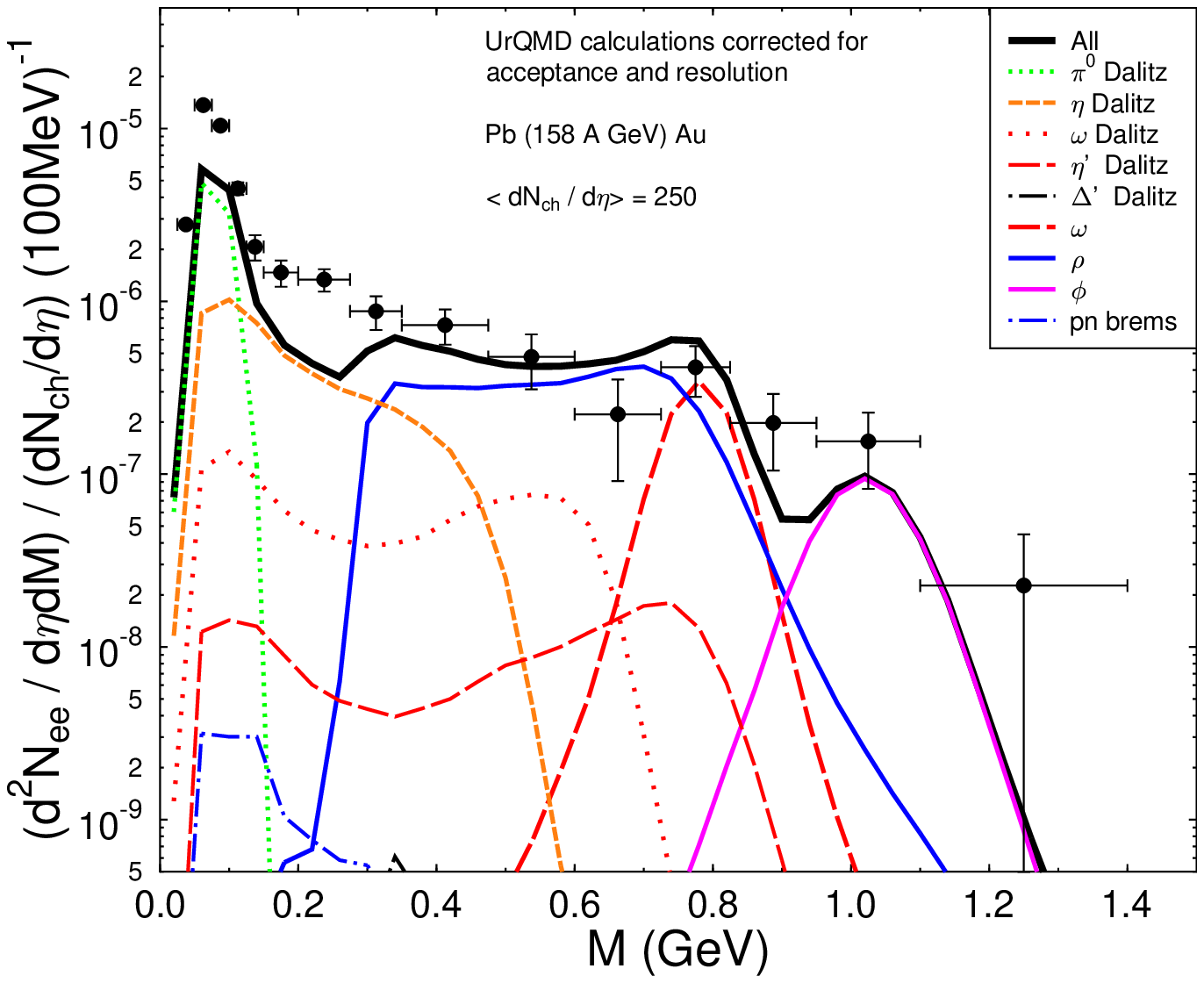}} \hfill
\parbox[b]{6cm}{
\caption{\small} \label{dilepton_pbau}
{\small Microscopic calculation of the dilepton 
production in the kinematic acceptance region of the CERES detector 
for $Pb+Au$ collisions at 158~$\rm{GeV}$. No in-medium effects are taken 
into account. Plotted data points are taken at CERES in '95.}
\vspace{2cm}
}}
\vspace{-0.5cm} 
\end{figure}

Fig.~\ref{dilepton_pbau} shows a microscopic UrQMD
calculation of the dilepton production in the kinematic 
acceptance region of the CERES detector for $Pb+Au$ collisions at 
158~$\rm{GeV}$. 
This is compared with the '95 CERES data\cite{agakishiev97b}. 
Aside from the difference at $M\approx 0.4$~$\rm{GeV}$ there is a strong 
enhancement at higher invariant masses. 
It is expected that this discrepancy at $m>1$~$\rm{GeV}$ could be filled up 
by direct dilepton production in meson-meson collisions\cite{liG98a} 
as well as by the mechanism of secondary Drell-Yan 
pair production proposed in\cite{spieles97a}.

\section{Strangeness production}

Strange particle yields are most interesting and useful probes to
examine excited nuclear matter 
\cite{raf8286,soff99plb,senger99,stock99,and98a,sikler99,cgreiner00} 
 and to detect the transition of
(confined) hadronic matter to quark-gluon-matter. The relative
enhancement of strange and especially multistrange particles in
central heavy ion collisions with respect to peripheral or proton
induced interactions have been suggested as a signature for the
transient existence of a QGP-phase \cite{raf8286}. Here the main idea is
that the strange (and antistrange) quarks are thought to be produced
more easily and hence also more abundantly in such a deconfined state
as compared to the production via highly threshold suppressed
inelastic hadronic collisions. 
The relative enhanement of (anti)hyperons
has clearly been measured by the WA97 an the NA49 collaboration in
Pb-Pb collisions as compared to p-Pb collisions
\cite{and98a,sikler99} 
. This
data has been investigated within microscopic transport 
models (e.g. UrQMD \cite{bass98}). In \cite{soff99plb,soff00} it was shown that
within such an appraoch strangeness enhancement is predicted for Pb-Pb
due to rescattering. However, for central Pb-Pb collisions the
experimentally observed hyperon yields are underestimated by the
calculation in \cite{soff99plb,soff00}. This result seems to confirm the
conclusion that a deconfined QGP is formed in Pb-Pb collisions at SPS.
But in
\cite{cgreiner00,cgreiner002} it was shown, that the antihyperon production  
by multi-mesonic reactions like 
$n_1 \pi + n_2 K \rightarrow \bar{Y} + p$ could drive these rare particles
towards local chemical quilibrium with pions, nucleons and kaons on a
timescale of 1-3 fm/c. Accordingly this mechanism, which is a
consequence of detailed balance could provide a
convenient explanation for the antihyperon yields at CERN-SPS energies
without any need of a deconfined quark-gluon-plasma phase. At the
moment such back-reactions cannot be handled within the present
transport codes. Therefore the aim for the future will be to find a
way to include these processes in microscopic transport models.
\section{Particle ratios}
Ideal gas model calculations have been used for a long time to
calculate particle production in relativistic heavy ion collisions, see e.g.
\cite{hahn86,hahn88,brau99,raf99,bec00,yen98}. 
Fitting the particle ratios as obtained 
from those ideal gas calculations to the experimental measured ratios
at  SIS, AGS and SPS for different energies and different colliding
systems yields a curve of  chemical freeze-out in the $T-\mu$ plane.
Now the question 
arises, how much the deduced temperature and chemical potentials 
depend on the model employed. Especially the influence of changing hadron
masses and effective potentials should be investigated, as has been
done for example in \cite{stoe78,thei83,scha91,springer}.
This is of special importance for the quest of a signal of the formation of a
 deconfined phase, i.e. the quark-gluon plasma. As deduced from lattice data 
\cite{kar98}, the critical temperature for the onset of a deconfined phase
 coincides with that of a chirally restored phase. Chiral effective models of
 QCD therefore can be utilized to give important insights on signals from a
 quark-gluon plasma formed in heavy-ion collisions.
  
\begin{figure}[h]
\centerline{\parbox[b]{6cm}{\epsfxsize=8cm
\vspace*{-2cm}
\epsfbox{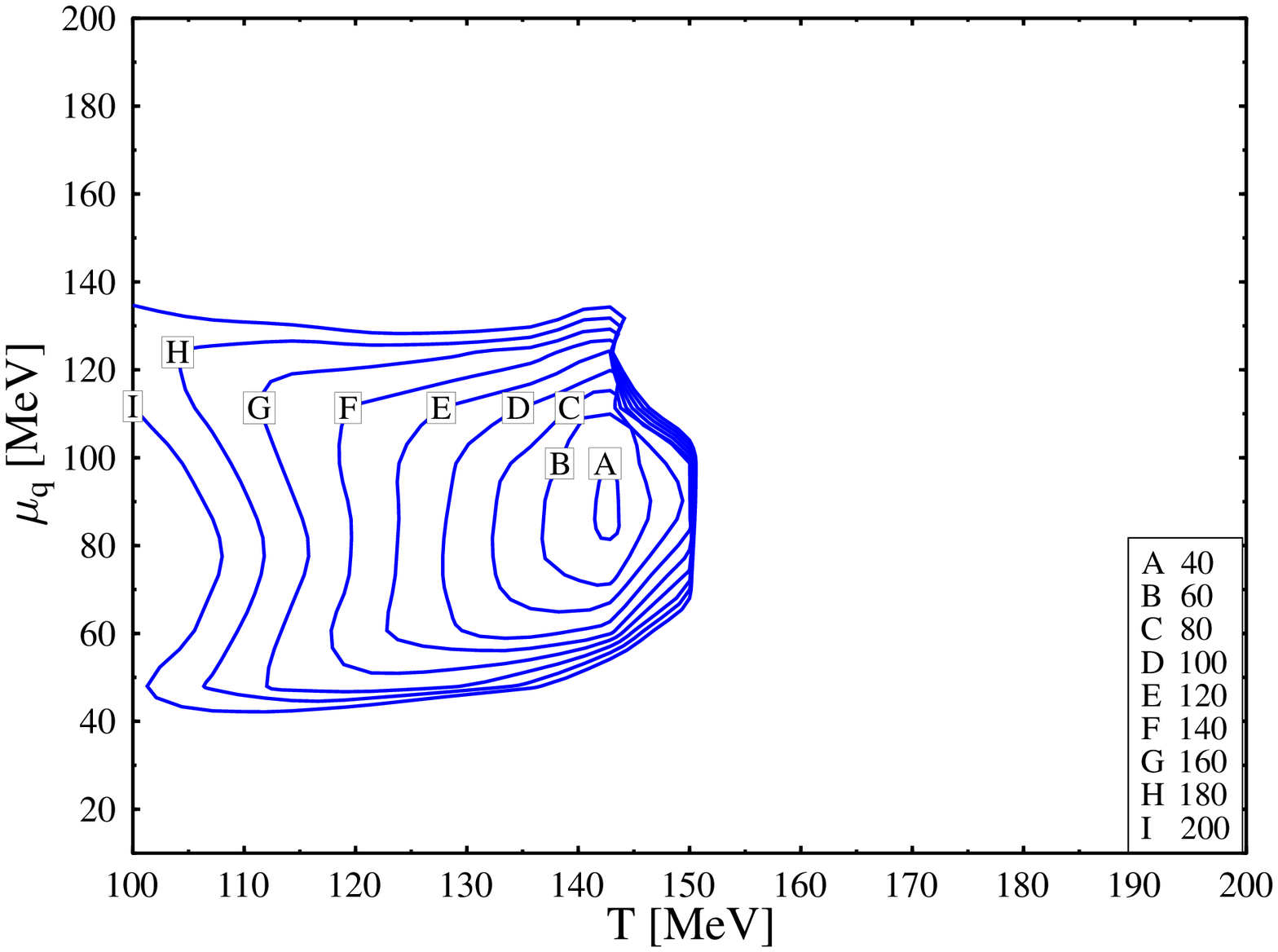}
\vspace{1cm}
}\hfill
\parbox[b]{6cm}{\epsfxsize=6cm
\vspace*{-.5cm}
\epsfbox{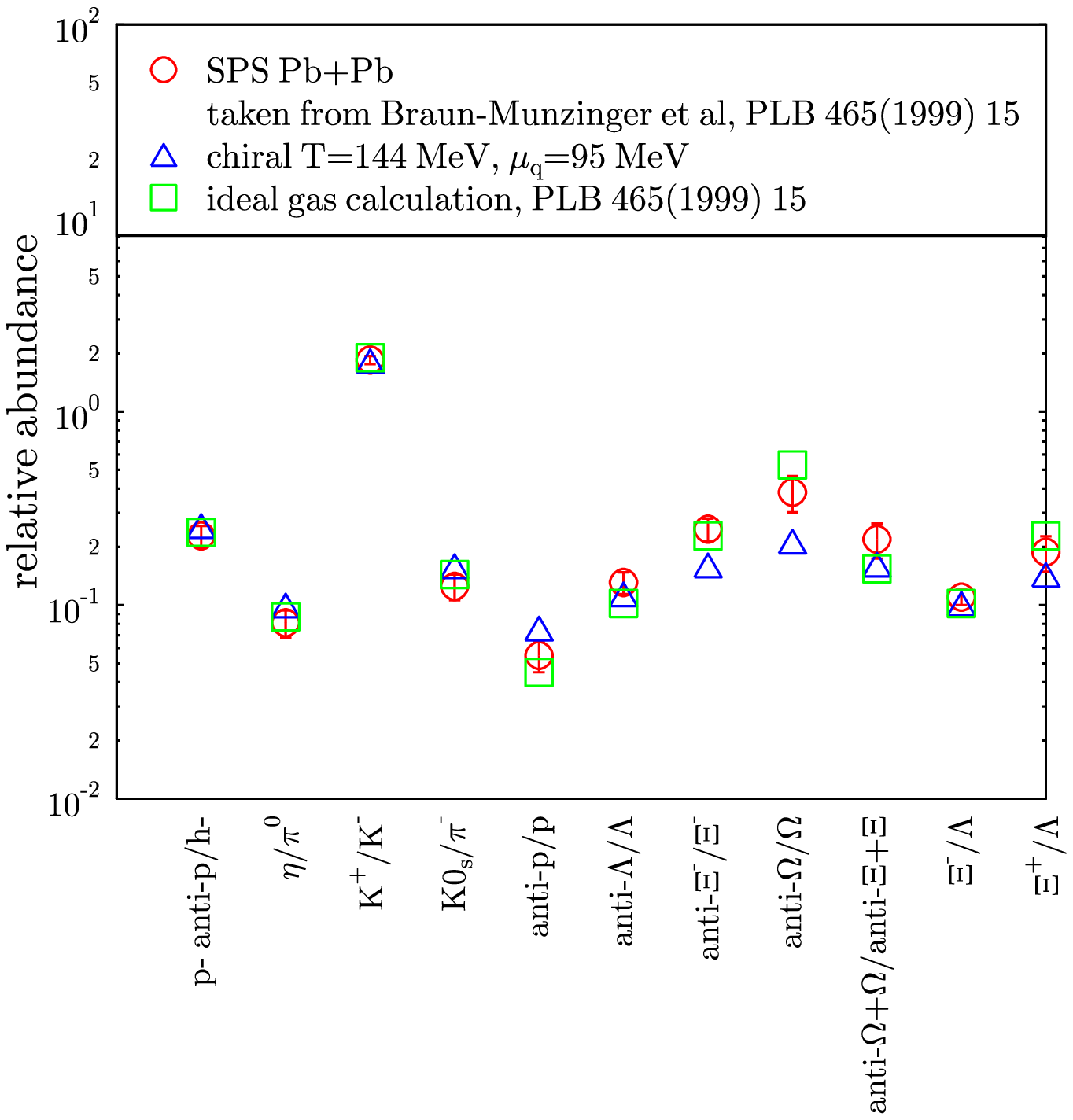}}\hfill
}
\vspace{-0.8cm}
{\caption{\small
$\chi^2$ (left) and resulting particle ratios compared to ideal gas
calculation and data (right) for chiral model, 
data taken from \protect\cite{brau99}. 
The best fit
parameters are $T = 144 \rm{MeV}$ and $\mu_q \approx 95 \rm{MeV}$.} 
\label{chi2chiral}}
\vspace*{-0.5cm}
\end{figure}

Therefore we compare experimental measurements for Pb+Pb collisions at
SPS with the ideal gas calculations and 
results obtained from a chiral SU(3) model \cite{springer,paper3}. 
This effective hadronic model 
predicts a chiral phase transition at $T \approx 150 \rm{MeV} $.
Furthermore the
model predicts changing hadronic masses and effective chemical
potentials, due to strong scalar and vector fields in hot and dense
hadronic matter, which are constrained by chiral symmetry from the 
QCD Lagrangean.\\
In \cite{brau99} 
the ideal gas model was fitted to particle ratios measured in Pb+Pb
collisions at SPS. The lowest 
$\chi^2$ is obtained for $T=168 \rm{MeV}$ and $\mu_q= 88.67 \rm{MeV}$. Using
these values as input for the chiral model leads to dramatic changes
due to the changing hadronic masses in hot and dense matter 
\cite{Zschiesche:2000zc}
and
therefore the freeze-out temperature and chemical potential have to be 
readjusted to account for the in-medium effects of the hadrons in the chiral 
model. We call the best fit the parameter set that gives a minimum in 
the value of $\chi^2$, with
$\chi^2 = \sum_i \frac {\left(r_i^{exp} - r_i^{model}\right)^2}{\sigma_i^2}.$
Here $r_i^{exp}$ is the experimental ratio, $r_i^{model}$ is the ratio
calculated in the model and $\sigma_i$ represents the error in the
experimental data points as quoted in \cite{brau99}.
The resulting values of $\chi^2$ for different $T-\mu$ pairs are shown 
in figure \ref{chi2chiral}.
 In all calculations $\mu_s$ was chosen
such that the overall net strangeness $f_s$ is zero. 
The best values for 
the parameters are $T = 144 \rm{MeV}$ and $\mu_q \approx 95
\rm{MeV}$. 
While the
value of the chemical potential does not change much compared to the
ideal gas calculation, the value of the temperature is lowered by
more than 20 $\rm{MeV}$. 
Furthermore Figure \ref{chi2chiral} shows,
that the dropping effective masses
and the reduction of the effective chemical potential make the
reproduction of experimentally measured particle ratios as seen at CERN's SPS 
within this
model impossible for $T > T_c$.
Using the best fit parameters  
a reasonable description of the particle ratios used in the fit
procedure can be obtained 
(see fig.\ref{chi2chiral}, data from \cite{brau99}).  

We want to emphasize, that in
spite of the strong assumption of thermal and chemical equilibrium
the obtained values for $T$ and $\mu$
differ significantly depending on the underlying model, i.e. whether and
how effective masses and effective chemical potentials are accounted for.
Note that we assume implicitly, that the particle ratios are 
determined by the medium effects and freeze out during the late stage 
expansion - no flavor changing collisions occur anymore, but the
hadrons can take the necessary energy to get onto their mass shall by
drawing energy from the fields.
Rescattering effects will alter our conclusion but are presumably small when 
the chemical potentials are frozen.      
\section{Collective flow and the EOS}
The in-plane flow has been proposed as a measure of the
''softening'' of the EoS~\cite{rischke95b}, therefore we investigate the
excitation function of directed in-plane flow. A three-fluid model
with dynamical unification of kinetically equilibrated fluid elements
is applied~\cite{hydro_review}.
This model assumes that a projectile- and a target fluid 
interpenetrate upon impact of the two nuclei, 
creating a third fluid  
via new source terms in the continuity equations 
for energy- and momentum flux. 
Those source terms are taken from energy- and rapidity 
loss measurements in high energy $pp$-collisions. 
The equation of state (EoS) of this model assumes equilibrium only 
in each fluid separately and allows for a first order phase transition 
to a quark gluon plasma in fluid 1, 2 or 3, 
if the energy density in the fluid under consideration 
exceeds the critical value for two phase coexistence. 
Pure QGP can also be formed in every fluid separately, 
if the energy density in that fluid exceeds the maximum energy 
density for the mixed phase.
Integrating up the collective momentum in $x$-direction at given
rapidity, and dividing by the net baryon number in that rapidity bin,
we obtain the so-called directed in-plane flow per nucleon.

\begin{figure}[h]
\vspace{-0.2cm}
\centerline{\parbox[b]{6cm}
{\epsfxsize=12pc
\epsfbox{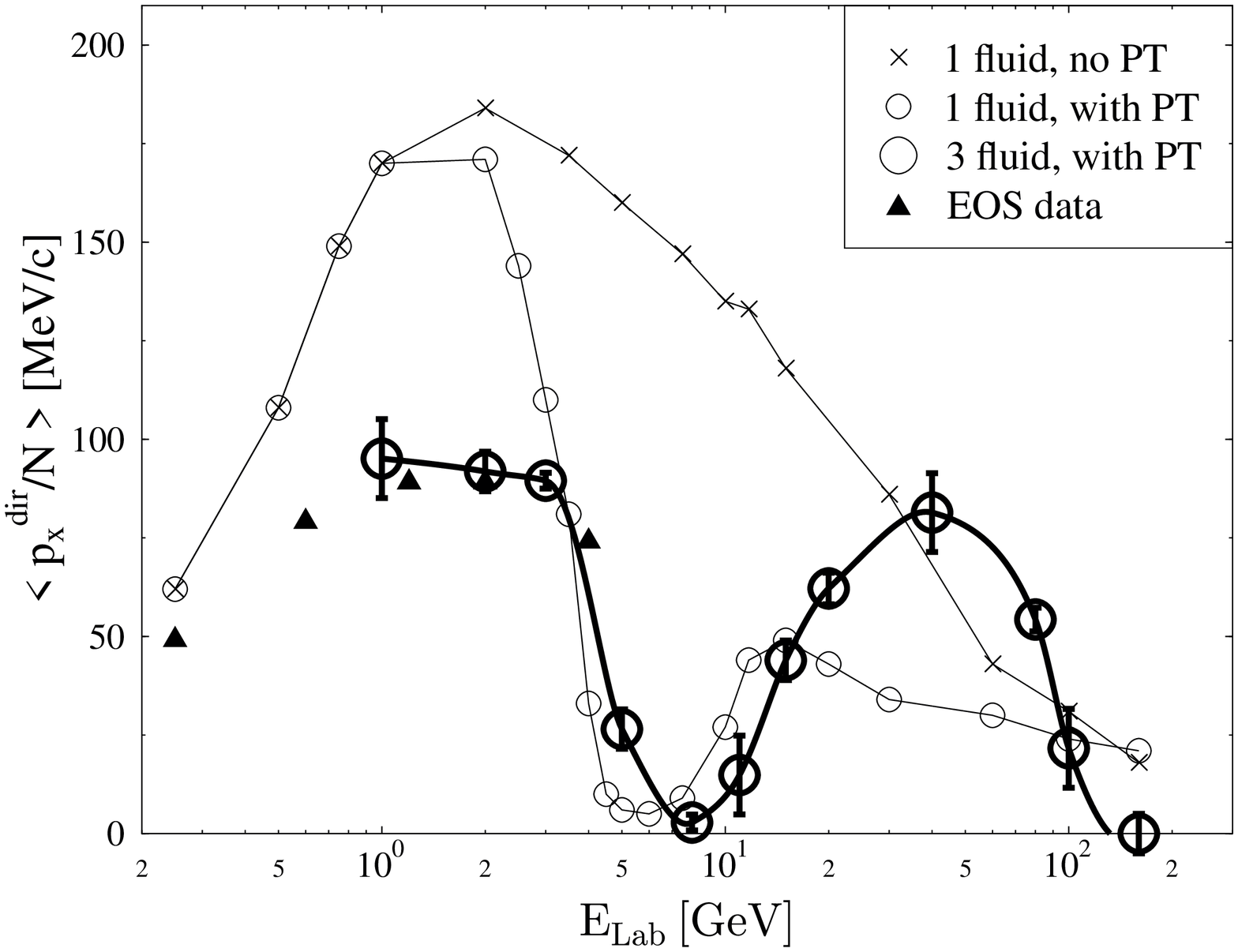}}
{\parbox[b]{6cm}
{\epsfxsize=14pc
\epsfbox{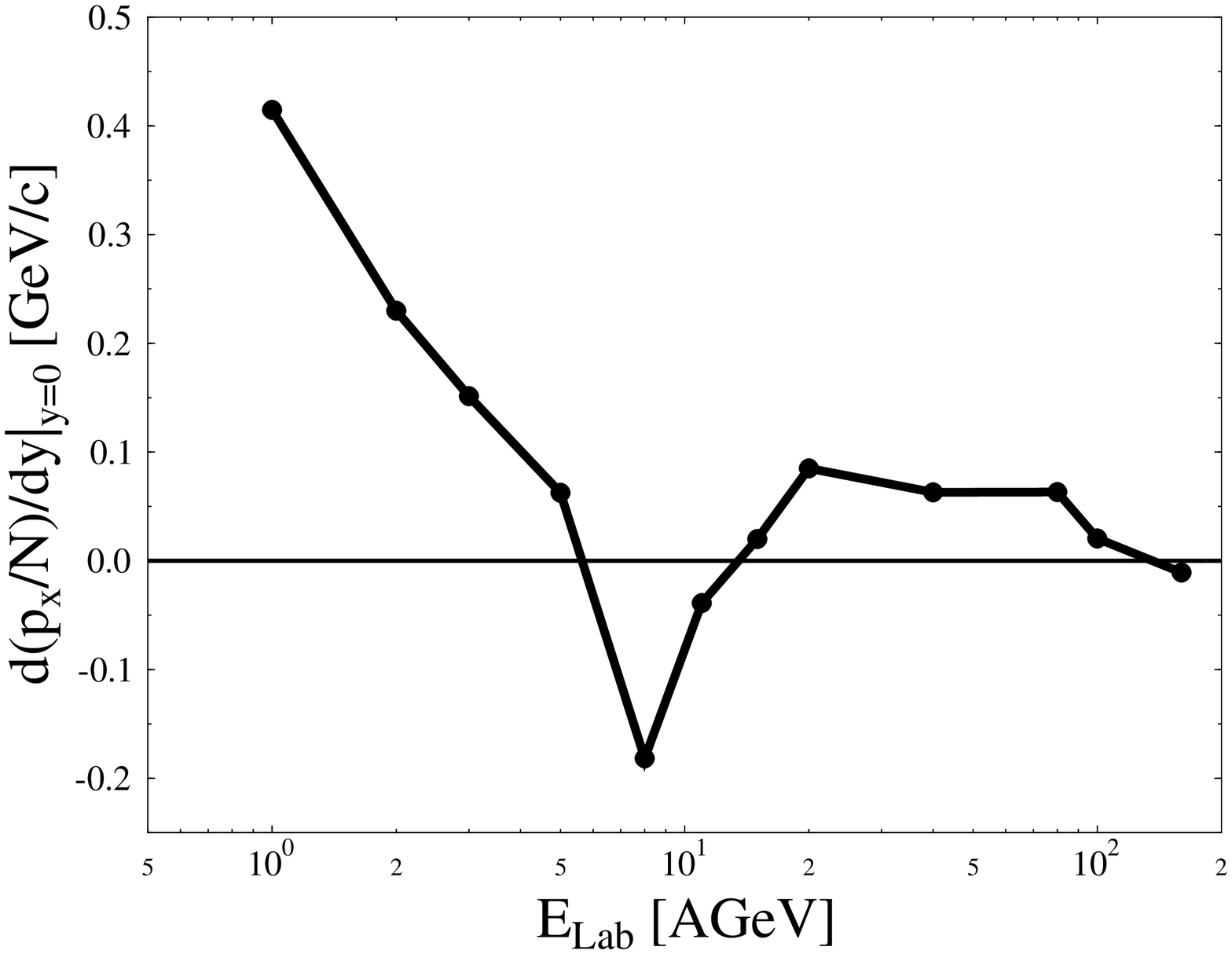}}}}
\vspace*{-0.5cm}
\caption{\label{hydro_flow}{\small}(Left) 
Excitation function of transverse flow as obtained from
three fluid hydrodynamics with a first order phase transition  
 and (Right) the slope of the directed in-plane momentum per nucleon at 
midrapidity.}
\vspace*{-0.5cm}
\end{figure}

Its excitation function (Fig.~\ref{hydro_flow}) shows a local 
minimum at 8 $\rm{AGeV}$ and rises until
a maximum around 40 $\rm{AGeV}$ is reached. 
Fig.~\ref{hydro_flow} shows the excitation function of directed flow
calculated in the three-fluid model in comparison to that obtained in a
one-fluid calculation. Due to non-equilibrium effects in the early stage of 
the reaction, which delay the  build-up of transverse pressure\cite{dirk}, 
the flow shifts to higher bombarding energies. 
While measurements of flow at AGS\cite{Liu} have found a decrease of directed flow with increasing bombarding
energy, a minimum has so far not been observed.

In a recent investigation of the directed flow excitation functions
\cite{ivanov} it has been shown, that the directed flow excitation
functions are sensitive to the underlying EoS and that a different EoS
can
predict a slowly and smoothly decrease of the averaged directed flow
as a function of bombarding energies. This different behaviour is due
to the different phase transitions in the underlying equations of
state. While in the two phase EoS based on a $\sigma-\omega$ model for
the hadronic phase and a bag model for the deconfined phase a
first-order phase transition occurs, the EoS in \cite{ivanov}
provides a continues phase transition of the cross-over type.

The
slope of the directed in-plane momentum per nucleon at midrapidity, 
$d(p_x/N)/dy$, is shown in Fig.~\ref{hydro_flow} as
a function of beam energy. We find a steady decrease of $d(p_x/N)/dy$
up to about top BNL-AGS energy, where the flow around midrapidity
even becomes negative due to preferred expansion towards
$p_x\cdot p_{long}<0$. At higher energy, $E_{Lab}\simeq40A$~$\rm{GeV}$,
the isentropic speed of sound becomes small and we encounter the following
expansion pattern : flow towards
$p_x\cdot p_{long}<0$ can not build up~! Consequently,
$d(p_x/N)/dy$ increases rapidly towards $E_{Lab}=20-40A$~$\rm{GeV}$,
decreasing again at even higher energy because of the more
forward-backward peaked kinematics which is unfavorable for
directed flow.

Thus, the $Pb + Pb$ collisions (40~$\rm{GeV}$) runs performed recently at the CERN-SPS
may provide a crucial test of the picture of a quasi-adiabatic
first-order hadronization phase transition
at small isentropic velocity of sound.

\section{Collective Flow at RHIC}
\begin{figure}[h]
\vspace*{-1cm}
\centerline
{\parbox[b]{6cm}{\epsfxsize=6.5cm
%\vspace*{-1.5cm}
\epsfbox{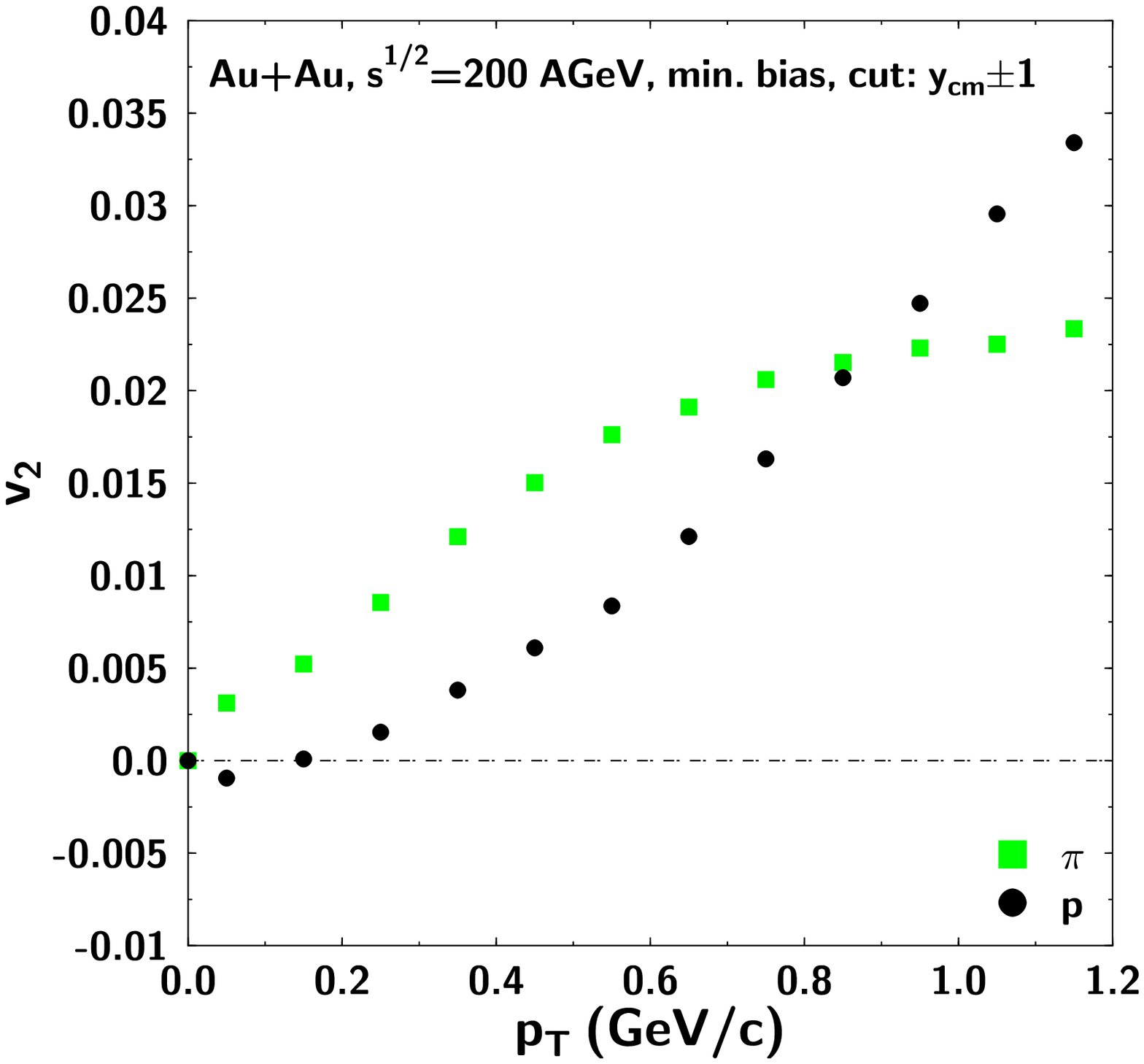}}\hfill
\parbox[b]{6cm}{\caption{\small} \label{v2}
{\small Elliptic flow parameter $v_2$ at midrapidity as a function of
transverse momentum in minimum biased Au+Au reactions 
at $\sqrt{s}=200 \rm{AGeV}$}
\vspace*{3cm}}
}
\vspace*{-0.5cm}
\end{figure}

Let us now compare the 
first results on elliptic flow ($v_2$) at $\sqrt{s_{NN}} = 130 \rm{GeV}$
as reported by the STAR-Collaboration   
\cite{star2000} with a string hadronic model simulation: 
The experimental data indicates a strongly rising $v_2$ as a function
of $p_t$ with an average $v_2$ value of 
$6 \%$ at midrapidity and $p_t$ approximately $600 \rm{MeV}$.
While the strong increase of $v_2$ with $p_t$ has been predicted by
the UrQMD model \cite{bleicher2000} the absolute magnitude of $v_2$ at
$p_t=600 \rm{MeV}$ is underpredicted by a factor 3
(cf. fig. \ref{v2}).

When the formation time of hadrons in the initial strings is strongly 
reduced (to mimic short mean free paths in the early interaction
region) the calculated flow values approach the hydrodynamic limit 
\cite{pasi,bleicher2000} 
and get in line with the measured elliptic flow
values. This shows, that
the pressure in the reaction zone is much higher than expected from
simple stringlike models 
and supports the breakdown of pure string hadronic dynamics
in the initial stage of Au-Au-collisions at RHIC energies. 
However, to get a consistent picture and to finally rule out the string
hadronic approach the $v_1$ values and transverse momentum
spectra \cite{Bleicher:2000pu} 
as given by the model calculation need to be exceeded by the
experimental data. 
\section{Insights from quark molecular dynamics}
Further insights about the possible formation of deconfined matter can be obtained from 
the Quark Molecular Dynamics Model (qMD)~\cite{Hofmann:1999jx} which explicitly includes quark 
degrees of freedom. The qMD can provide us with detailed information about the dynamics of the quark 
system and the parton-hadron conversion. Correlations between the quarks clustering to build new 
hadrons can be studied~\cite{Scherer:2000aa}. 

\begin{figure}[h]
\vspace*{-0.5cm}
\centerline
{\parbox[b]{6cm}{\epsfxsize=6.5cm \vspace*{-0.2cm}\epsfbox{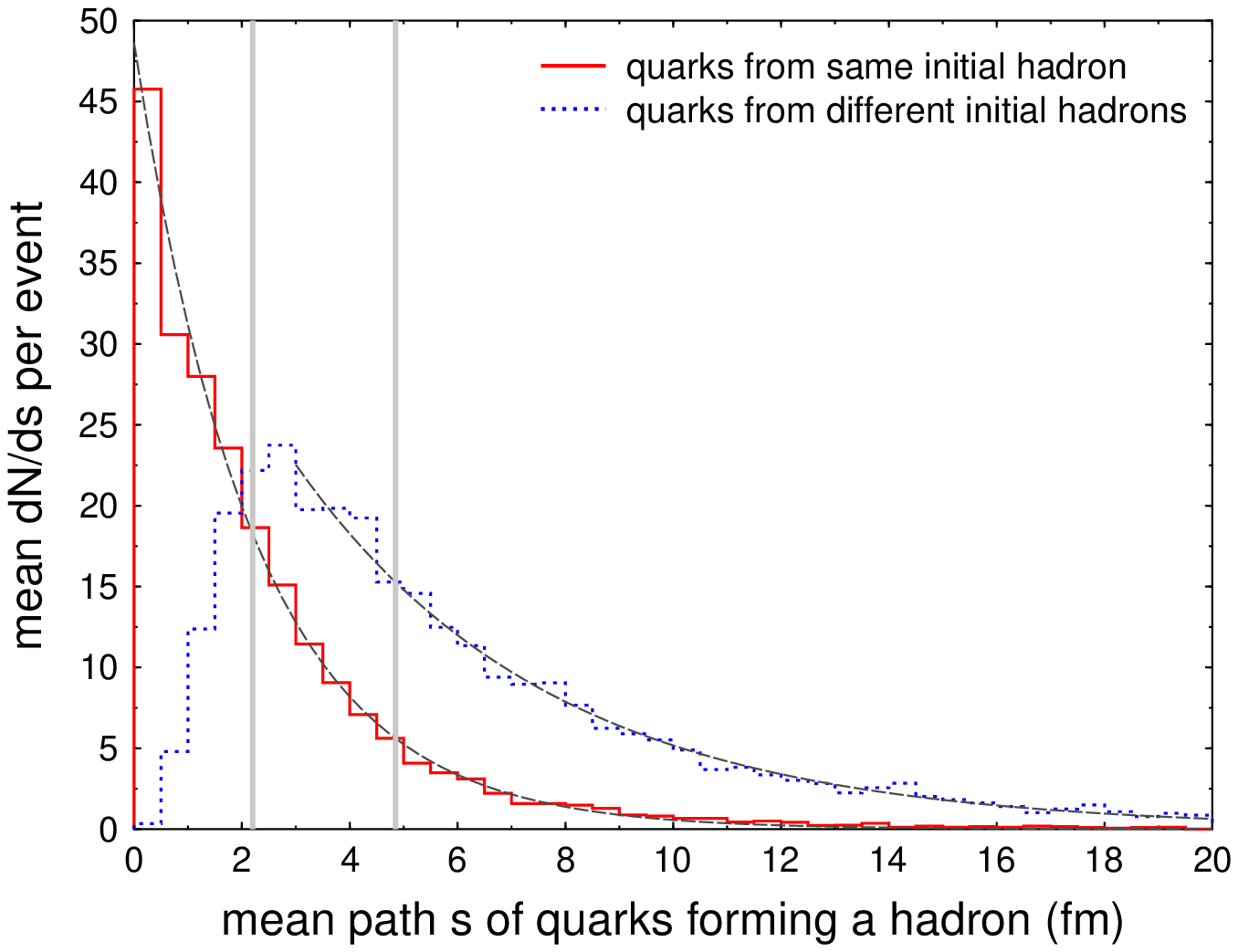}}\hfill
\parbox[b]{6cm}{\caption{\small} \label{qmd1}
{\small Hadronization in S+Au collisions at SPS (200 GeV/$N$): Number density distribution 
of mean diffusion path of quarks forming a hadron from the same initial hadron (solid line) 
and from different initial hadrons (dashed line) within qMD. Fitting the decay profiles yields
diffusion lengths of $2.2\,\mathrm{fm}$ and $4.8\,\mathrm{fm}$, respectively.}
\vspace*{1.3cm}}
}
\vspace*{-0.5cm}
\end{figure}

Figure \ref{qmd1} shows (for S+Au collisions at SPS energies of $200\,\mathrm{GeV}/N$)
the number distribution for the mean path travelled by quarks forming a hadron (a) from 
the same initial hadron (solid line) and (b) from different initial hadrons (dotted line). 

A measure of the relative mixing within the quark system and thus for thermalization
is the relative number of hadrons formed by quarks from the same initial hadron correlation 
versus hadrons formed by quarks from different initial hadron correlations. This ratio is $r = 0.574 \pm 0.008$ 
for the S+Au collision. Since a value of $r=1$ would indicate 
complete rearrangement of quarks and thus complete loss of correlations in the quark system, one would expect 
a much larger value of $r$, considering the presumed transition to the quark-gluon plasma in Pb+Pb
collisions at $160\,\mathrm{GeV}/N$,

\section*{Outlook}

The latest data of CERN/SPS on flow, electro-magnetic probes, 
strange particle yields (most importantly multistrange (anti-)hyperons) 
and heavy quarkonia will be interesting to follow closely. 
Simple energy densities estimated from rapidity distributions 
and temperatures extracted from particle spectra indicate 
that initial conditions could be near or just 
above the domain of deconfinement and chiral symmetry restoration.
Still the quest for an {\em unambiguous} signature remains open.

Directed flow has been discovered 
-- now a flow excitation function, filling the gap between 10 $\rm{AGeV}$ (AGS) 
and 160 $\rm{AGeV}$ (SPS), will be extremely interesting: look for the 
softening of the QCD equation of state in the coexistence region. 
The investigation of the physics of high baryon density 
(e.g. partial restoration of chiral symmetry via properties 
of vector mesons) has been pushed forward by the 40 $\rm{GeV}$ run at SPS.
Also the excitation function of particle yield 
ratios ($\pi/{\rm p}, {\rm d}/{\rm p}, {\rm K}/\pi ...$) and, in particular, 
multistrange (anti-)hyperon yields, can be a sensitive probe of 
physics changes in the EoS.
The search for novel, unexpected forms of matter, e.g. {\em hypermatter}, 
{\em strangelets} or even {\em charmlets} is intriguing. Such exotic QCD 
multi-meson and multi-baryon configurations would extend 
the present periodic table of elements into hitherto 
unexplored dimensions. A strong
experimental effort should continue in that direction.

Now we have entered the exciting RHIC era,
where the predicted deconfined and chirally restored
phase should be formed and live long enough to 
produce clear and unambigious signals
of it's existence. The LHC-program will top this scientific endeavour
in 4 years.

\section*{Acknowledgments}

This work was supported by DFG, GSI, BMBF, Graduiertenkolleg Theoretische 
und Experimentelle Schwerionenphysik, the A.~v.~Humboldt Foundation, 
and the J.~Buchmann Foundation.


\begin{thebibliography}{99}

\bibitem{MaT86}
T. Matsui and H. Satz, Phys. Lett. {\bf B178},  416  (1986).

\bibitem{KhD96}
D. Kharzeev, Nucl. Phys. {\bf A610},  418c  (1996).

\bibitem{ger}
L. Gerland, L. Frankfurt, M. Strikman, H. St\"ocker, and W. Greiner, Phys. Rev.
  Lett. {\bf 81},  762  (1998).

\bibitem{slac}
R.~L. Anderson {\it et~al.}, Phys. Rev. Lett. {\bf 38},  263  (1977).

\bibitem{farrar}
G.~R. Farrar, L.~L. Frankfurt, M.~I. Strikman, and H. Liu, Phys. Rev. Lett.
  {\bf 64},  2996  (1990).

\bibitem{werner}
L. Frankfurt, W. Koepf, and M. Strikman, Phys. Rev. {\bf D54},  3194  (1996).

\bibitem{kharzeev}
D. Kharzeev, C. Lourenco, M. Nardi, and H. Satz, Z. Phys. {\bf C74},  307
  (1997).

\bibitem{hufner}
C. Gerschel and J. Hufner, Phys. Lett. {\bf B207},  253  (1988).

\bibitem{carlos}
C. Lourenco, Nucl. Phys. {\bf A610},  552c  (1996).

\bibitem{bass98}
S.~A. Bass {\it et~al.}, Prog. Part. Nucl. Phys. {\bf 41},  225  (1998).

\bibitem{spieles99}
C.~Spieles, R.~Vogt, L.~Gerland, S.~A.~Bass, M.~Bleicher, H.~Stocker and W.~Greiner,
%``Modelling J/Psi production and absorption in a microscopic  nonequilibrium approach,''
Phys.\ Rev.\  {\bf C60} (1999) 054901
[hep-ph/9902337].

\bibitem{romana}
A.Romana {\it et~al.}, in Proceedings of the XXXIIIrd Rencontres de Moriond,
  March 1998, Les Arcs, France.

\bibitem{na50neu}
M.~C. Abreu {\it et~al.}, Phys. Lett. {\bf B477},  28  (2000).

\bibitem{reviewvogt}
R. Vogt, Phys. Rept. {\bf 310},  197  (1999).

\bibitem{cap00}
A. Capella, E.~G. Ferreiro, and A.~B. Kaidalov, hep-ph/0002300  (2000).

%-------------------------------------------

\bibitem{koch97}
	V.~Koch, Int.~Jour.~Mod.~Phys.\ {\bf E6} (1997) 203.

\bibitem{cassing97a}
	W.~Cassing, E.~L.~Bratkovskaya, R.~Rapp, and J.~Wambach,
	Phys.~Rev.\ C{\bf 57} (1998) 916

\bibitem{agakishiev97b}
	G.~Agakishiev et al., Phys.~Lett.\ {\bf B402} (1998) 405.

\bibitem{liG98a}
	G.~Q.~Li and C.~Gale, Phys.~Rev.~{\bf C58} (1998) 2914.

\bibitem{spieles97a}
	C.~Spieles et al., Eur.~Phys.~J. {\bf C5} (1998) 349

%-------------------------------------------
\bibitem{raf8286}
J.~Rafelski, B.~M\"uller 
\PRL {\bf 48}, (1982) 1066; (E) {\bf 56} (1986) 2334;
P.~Koch, B.~M\"uller, J.~Rafelski 
{\it Phys.~Rep.} {\bf 142}, (1986) 167;
P.~Koch, B.~M\"uller, H.~St\"ocker, W.~Greiner 
{\it Mod.~Phys.~Lett.}~{\bf A3}, (1988) 737

\bibitem{soff99plb}
S.~Soff, S.~A.~Bass, M.~Bleicher, L.~Bravina, M.~Gorenstein,   
E.~Zabrodin, H.~St\"ocker, W.~Greiner \,
\PL~{\bf B471}, (1999) 89 and refs. therein

\bibitem{senger99}
P.~Senger, H.~Str\"obele 
\JPG {\bf 25}, (1999) R59 

\bibitem{stock99}
R.~Stock 
\PL {\bf B456}, (1999) 277

\bibitem{and98a}
E.~Andersen \etal (WA97 collaboration)
\PL~{\bf B433}, (1998) 209;
S.~Margetis \etal (NA49 collaboration)
\JPG~{\bf 25}, (1999) 189 

\bibitem{sikler99}
F.~Sikler \etal (NA49 collaboration)
\NP~{\bf A661}, (1999) 

\bibitem{cgreiner00}
C.~Greiner, S.~Leupold  nucl-th/0009036


%\cite{Soff:2000ae}
\bibitem{soff00}
S.~Soff {\it et al.},
%``Enhanced strange particle yields: Signal of a phase of massless  particles?,''
\JPG ~in print, nucl-th/0010103.
%%CITATION = NUCL-TH 0010103;%%

%\cite{Greiner:2000vb}
\bibitem{cgreiner002}
C.~Greiner,
%``Do chemically saturated antihyperon abundancies signal the quark gluon  plasma?,''
%in {\it NONE}
nucl-th/0011026.
%%CITATION = NUCL-TH 0011026;%%

%------------------------------------------

\bibitem{hahn86}
D. Hahn and H. St{\"o}cker, Nucl. Phys. {\bf A452},  723  (1986).

\bibitem{hahn88}
D. Hahn and H. St{\"o}cker, Nucl. Phys. {\bf A476},  718  (1988).

\bibitem{brau99}
P. Braun-Munzinger, J. Heppe, and J. Stachel, Phys. Lett. B {\bf 465},  15
  (1999).

\bibitem{raf99}
J. Rafelski and J. Letessier, nucl-th/9903018  (1999).

\bibitem{bec00}
F. Becattini, J. Cleymans, A. Keranen, E. Suhonen, and K. Redlich,
  hep-ph/0002267  (2000).

\bibitem{yen98}
G.~D. Yen and M.~I. Gorenstein, Phys. Rev. {\bf C59},  2788  (1999).

\bibitem{stoe78}
H. St{\"o}cker and W. Greiner, Z. Phys. A {\bf 286},  121  (1978).

%\cite{Thei83}
\bibitem{thei83}
J.~Theis, G.~Graebner, G.~Buchwald, J.~A.~Maruhn, W.~Greiner, H.~St\"ocker and J.~Polonyi,
%``Phase Transition Of The Nucleon - Anti-Nucleon Plasma In A Relativistic Mean Field Theory,''
Phys.\ Rev.\  {\bf D28} (1983) 2286.
%%CITATION = PHRVA,D28,2286;%%

\bibitem{scha91}
J. Schaffner, I.~N. Mishustin, L.~M. Satarov, H. St{\"o}cker, and W. Greiner,
  Z. Phys. {\bf A341},  47  (1991).

\bibitem{springer}
D. Zschiesche, P. Papazoglou, S. Schramm, C. Beckmann, J. Schaffner-Bielich, H.
  St{\"o}cker, and W. Greiner, Springer Tracts in Modern Physics {\bf 163},
  129  (2000).

\bibitem{kar98}
F. Karsch, hep-lat/9903031  (1998).

\bibitem{paper3}
P. Papazoglou, D. Zschiesche, S. Schramm, J. Schaffner-Bielich, H. St{\"o}cker,
  and W. Greiner, Phys. Rev. C {\bf 59},  411  (1999).
%\cite{Zschiesche:2000zc}
\bibitem{Zschiesche:2000zc}
D.~Zschiesche, L.~Gerland, S.~Schramm, J.~Schaffner-Bielich, H.~St\"ocker and W.~Greiner,
%``Critical review of quark gluon plasma signals,''
%in {\it NONE}
nucl-th/0007033.
%%CITATION = NUCL-TH 0007033;%%

%
%------------------------------------------------------------------
%
\bibitem{rischke95b}
	D.~H.~Rischke, Y.~P\"urs\"un, J.A.~Maruhn, H.~St\"ocker, W.~Greiner,
	Heavy~Ion~Physics~{\bf 1}~(1995)~309.

\bibitem{hydro_review}
	J.~Brachmann, A.~Dumitru, J.A.~Maruhn, H.~St\"ocker, W.~Greiner, D.H.~Rischke,
	Nucl.~Phys.~{\bf A619}~(1997)~391. 
\bibitem{dirk}
	H.~Sorge, Phys. Rev. Lett. {\bf 78}, 2309 (1997) 
\bibitem{Liu}
	H.~Liu et al. (E895 Collaboration), Nucl. Phys. {\bf A638}, 451c (1998)
%\cite{Ivanov:2000dr}
\bibitem{ivanov}
Y.~B.~Ivanov, E.~G.~Nikonov, W.~Noerenberg, A.~A.~Shanenko and V.~D.~Toneev,
%``Directed flow of baryons in heavy-ion collisions,''
nucl-th/0011004.
%%CITATION = NUCL-TH 0011004;%%

%----------------------------------------------------
%\cite{Ackermann:2000tr}
\bibitem{star2000}
K.~H.~Ackermann {\it et al.}  [STAR Collaboration],
%``Elliptic flow in Au + Au collisions at s(N N)**(1/2) = 130-GeV,''
nucl-ex/0009011.
%%CITATION = NUCL-EX 0009011;%%	

%\cite{Bleicher2000sx}
\bibitem{bleicher2000}
M.~Bleicher and H.~St\"ocker,
%``Anisotropic flow in ultra-relativistic heavy ion collisions,''
hep-ph/0006147.
%%CITATION = HEP-PH 0006147;%%

\bibitem{pasi}
P. Huovinen, priv. comm.;\\
P.~F.~Kolb, J.~Sollfrank and U.~Heinz,
%``Elliptic and hexadecupole flow from AGS to LHC energies,''
Phys.\ Lett.\  {\bf B459} (1999) 667
[nucl-th/9906003].
%%CITATION = NUCL-TH 9906003;%%
%\href{http://www.slac.stanford.edu/spires/find/hep/www?eprint=NUCL-TH/9906003}{SPIRES}.

%\cite{Bleicher:2000pu}
\bibitem{Bleicher:2000pu}
M.~Bleicher {\it et al.},
%``Global observables and secondary interactions in central Au + Au  reactions at s**(1/2) = 200-A-GeV,''
Phys.\ Rev.\  {\bf C62} (2000) 024904
[hep-ph/9911420].
%%CITATION = HEP-PH 9911420;%%

%--------------------------------------------------

\bibitem{Hofmann:1999jx}
M.~Hofmann, S.~Scherer, M.~Bleicher, L.~Neise, H.~St\"ocker, and W.~Greiner,
Phys.\ Lett.\ {\bf B478} (200) 161 

\bibitem{Scherer:2000aa}
S.~Scherer, M.~Hofmann, M.~Bleicher, L.~Neise, H.~St\"ocker, and W.~Greiner,
N.\ Journ.\. Phys.\  {\em to be publ.\ }





\end{thebibliography}
\end{document}